\documentclass[a4paper]{article}
\newcommand{\keywords}[1]{\textbf{Keywords:}\quad #1}
 
\usepackage{graphics}
\usepackage{adjustbox,lipsum}
\usepackage[english]{babel}
\usepackage[utf8]{inputenc}
\usepackage{amsmath}
\usepackage{graphicx}
\usepackage{hyperref}
\usepackage{url}
\usepackage{subcaption}
\usepackage[colorinlistoftodos]{todonotes}
\usepackage{ upgreek }
\usepackage{amsfonts}
\usepackage{caption}
\usepackage{float}
\usepackage{subcaption}
\usepackage{multirow}
\usepackage{enumitem}
\usepackage{booktabs}

\title{Bayesian Non-parametric Simultaneous Quantile Regression for Complete and Grid Data}

\author{Priyam Das\footnote{Deaprtment of Statistics, North Carolina State University, NC, USA} $\/$ and Subhashis Ghosal \footnote{Deaprtment of Statistics, North Carolina State University, NC, USA}}

\date{\today}

\begin{document}
\maketitle
\begin{abstract}
In this paper, we consider Bayesian methods for non-parametric quantile regressions with multiple continuous predictors ranging values in the unit interval. In the first method, the quantile function is assumed to be smooth over the explanatory variable and is expanded in tensor product of B-spline basis functions. While in the second method, the distribution function is assumed to be smooth over the explanatory variable and is expanded in tensor product of B-spline basis functions. Unlike other existing methods of non-parametric quantile regressions, the proposed methods estimate the whole quantile function instead of estimating on a grid of quantiles. Priors on the B-spline coefficients are put in such a way that the monotonicity of the estimated quantile levels are maintained unlike local polynomial quantile regression methods. The proposed methods have also been modified for quantile grid data where only the percentile range of each response observations are known. Simulations studies have been provided for both complete and quantile grid data. The proposed method has been used to estimate the quantiles of US household income data and North Atlantic hurricane intensity data. 
\end{abstract}

\section{Introduction} 
Quantile regression is one of a popular alternative to mean regression when the data is non-normal,skewed or heteroscedastic. Quantile regression is also useful when our objective is to make inferences about the population at different quantile levels. Linear quantile regression \hspace{.2cm} was \hspace{.2cm} first proposed in  \cite{Koenkar1978}. Various other frequentist methods for quantile regression can be found in \cite{Koenkar2005}. \cite{Yu2001} first introduced \hspace{.4cm} the \hspace{.4cm} quantile \hspace{.4cm} regression \hspace{.4cm} using \hspace{.4cm} Bayesian \hspace{.4cm} methods. \hspace{.4cm} Later \hspace{.5cm} \cite{Kottas2001}, \hspace{.5cm}\cite{Gelfand2003}, \newline \cite{Kottas2009}, \cite{Geraci2007} proposed a few methods on generalization and extension of single level quantile regression under different possible scenarios.

 The main disadvantage of considering separate quantile regression using single level quantile regression is that the natural ordering among different quantiles can not be ensured. Addressing the non-crossing issue, \cite{He1997} proposed a quantile regression method assuming the response variable to be heteroskedastic. \cite{Neocleous2008} proposed a method to estimate the quantile curve using linear interpolation from a estimated gird of quantile curves. \cite{Takeuchi2004} and \cite{Takeuchi2006} proposed non-crossing quantile regression methods using support vector machine (SVM, \cite{Vapnik1995}). Later \cite{Shim2009} used doubly penalized kernal machine (DPKM) for estimating non-crossing quantile curves. 
 
 \cite{Dunson2005} and \cite{Liu2011} proposed quantile regression methods for a grid of quantiles addressing the monotonicity constraint. Later  \hspace{.3cm} \cite{Wu2009}, \hspace{.3cm} \cite{Reich2012}, \hspace{.3cm} \cite{Reich2011}, \newline \cite{Reich2013} proposed linear quantile regression methods addressing the \hspace{.1cm} non-crossing \hspace{.2cm} issues. \hspace{.2cm} Recently, \hspace{.2cm} \cite{Tokdar2012} \hspace{.2cm} and \newline \cite{DasQR_1_2016} proposed simultaneous linear quantile regression methods for univariate explanatory variables. \cite{Yang2016} extended that simultaneous linear quantile regression method to handle multivariate predictor case. \cite{DasQR_2_2016} also extended the method proposed in \cite{DasQR_1_2016} in the spatio-temporal simultaneous quantile regression context where the response variable is assumed to be linearly dependent on one of the explanatory variables and varying non-parametrically with rest of the explanatory variables.

One of the shortcomings of using linear quantile regression method is that it is not able to reveal a higher degree polynomial trend in the quantile curves. For example, while regressing household income data of a country over time, lower quantile levels maybe linear with time but the upper quantile levels may evolve very differently, for instance lower quantile level may increase linearly with time but upper quantile level may increase faster. In that case, using linear quantile regression will not be appropriate since it would only give the closest possible linear approximation of the quantile curves. Quantile regression methods addressing this issue were proposed in \cite{Chaudhuri1991}, \cite{Chaudhuri1991a}. \cite{Yu1998} proposed the local linear quantile regression method based on the techniques proposed in \cite{Fan1994},\hspace{.2cm}  \cite{Fan1996}, \hspace{.2cm}\cite{Chaudhuri1991}.\hspace{.2cm} \cite{Chaudhuri2002} \hspace{.2cm}and \cite{Honda2010} proposed alternative methods of quantile regressions. A few other \hspace{.2cm} non-parametric \hspace{.2cm} quantile \hspace{.2cm} regression \hspace{.2cm}  methods \hspace{.2cm} were proposed in \hspace{.5cm} \cite{Samanta1989}, \hspace{.5cm} \cite{Li2008} and \cite{Gannoun2002}. \cite{Koenker2015} provided a detailed description of local linear quantile regression method using software R. Alongside, he also proposed an alternative method called local spline quantile regression.

As long as non-linear quantile regression is concerned, most of the above-mentioned methods do not take care of the non-crossing issues. Although, \cite{Bondell2010} proposed a non-parametric quantile regression method addressing the non-crossing issue, it can only estimate a set of non-crossing curves for a given grid of quantiles and it does not estimate the whole quantile regression function. Secondly, the quantile regression estimates obtained by this method is sensitive to the number and location of chosen quantile grids. Instead of estimating the quantile curves at a given set of quantiles, it is more desirable to estimate the whole quantile curve simultaneously to emerge the broader picture.

In this paper, we propose two Bayesian methods for quantile regression using B-spline. In the first method, the entire quantile function is modeled by a B-spline series expansion. For each of the explanatory variables, a corresponding B-spline basis function is considered. The whole quantile function is obtained via tensor product of B-spline basis functions corresponding to each dimension of explanatory variables and one corresponding to the quantile level. The prior on the B-spline coefficients is put in such a way that the monotonicity of the quantile curves is maintained. We name this method `Non-parametric Simultaneous Quantile Regression (NPSQR)'. In the second method, instead of the quantile function, the conditional distribution function is estimated non-parametrically using B-spline basis expansion. Similar to NPSQR, in this method also, corresponding to each dimension of the explanatory variable, a B-spline basis is considered. Again, the prior of the coefficients of the B-spline basis functions is put in such a way that the monotonicity of the distribution function is maintained. In this case also the whole distribution function is given by tensor product of the B-spline basis functions. The conditional distribution function is inverted to obtain the quantile regression function. The use of splines, which are piece-wise polynomials, allow efficient inversion through a combination of analytical and numerical technique. We name this method to be `Non-parametric Distribution Function Simultaneous Quantile Regression (NPDFSQR)'. Further using both of these two approaches, we propose the method of estimating the quantile curves when only the data with frequencies of the observations in each quantile range are available. 

\section{Proposed Bayesian Method}
\label{bayesian_model_here}
Suppose $\{(X_{1i}, \ldots, X_{di})\}_{i=1}^n$ and $\{Y_i\}_{i=1}^n$ denote the $d$-dimensional explanatory variable and the response variable respectively. Using monotonic transformation, each coordinates of the explanatory variable and the response variable are transformed into unit interval. 

\subsection{Non-parametric Modeling of Quantile Function}
\label{npsqr_model}
Suppose $Q(\tau|\mathbf{x})$ denotes the conditional quantile function of $Y$ given $\mathbf{X} = \mathbf{x}=(x_1,\ldots,x_d)$. 
A B-spline function of degree $m_1$ (i.e., degree of piece-wise polynomial is $m$) with knot sequence $0=t_0<t_1<\cdots<t_{p_1}=1$ has $(p_1+m_1)$ basis functions. Let $\{B_{j,m_1}(\cdot)\}_{j=1}^{p_1+m_1}$ denote B-spline basis functions of degree $m_1$ on the above-mentioned knot sequence. For simplicity we consider equidistant knots, i.e., $(t_i-t_{i-1}) = 1/p_1$ for $i=1,\ldots,p_1$. Hence the quantile function is given by
\begin{align}
Q(\tau|\mathbf{x}) = \sum_{j=1}^{p_1+m_1}\theta_j(\mathbf{x})B_{j,m_1}(\tau), \; 0=\theta_1(\mathbf{x})< \cdots < \theta_{p_1+m_1}(\mathbf{x})=1
\end{align}
where $\theta_j(\mathbf{x}), j=1,\ldots, p_1+m_1$, are the coefficients of B-splines basis expansion of $Q(\tau|\mathbf{x})$. Thus in the above-mentioned equation it is noted that the coefficients of the basis functions used to expand the quantile function are dependent on the explanatory variable $\mathbf{X}$. Secondly, the imposed monotonicity condition on the B-spline coefficients $\{\theta_j(\mathbf{x})\}_{j=1}^{p_1+m_1}$ ensures the monotoniticy of the quantile levels (\cite{Boor2001}).

Now to put a prior, the functions $\{\theta_j(\mathbf{x})\}_{j=1}^{p_1+m_1}$, it is expanded using $d$-dimensional tensor product of the B-spline basis functions of degree $m_2$. We use the knot sequence $\{s_i\}_{i=1}^{p_2}$ such that $0=s_0<s_1<\cdots<s_{p_2}=1, (s_i-s_{i-1}) = 1/p_2 \; \text{for} \; i=1,\ldots,p_2$ for all the coordinates of the explanatory variable $\mathbf{X}$. Then $\theta_j(\mathbf{x})$ is given by
\begin{align}
\theta_j(\mathbf{x}) = \sum_{k_1=1}^{p_2+m_2} \cdots \sum_{k_d=1}^{p_2+m_2}\alpha_{jk_1 \cdots k_d}B_{k_1,m_2}(x_1)\ldots B_{k_d,m_2}(x_d).
\end{align}
Then the parameters which need to be estimated are given by
\begin{align}
0=\alpha_{1k_1\cdots k_d}<\cdots<\alpha_{(p_1+m_1)k_1\cdots k_d}=1,\; (k_1,\ldots,k_d) \in \{1,\ldots,(p_2+m_2)\}^d.
\label{npsqr_parameter}
\end{align}

\subsection{Non-parametric Modeling of Distribution Function}
\label{npdfsqr_model}
For modeling the distribution function with B-spline basis functions, we adopt a similar technique. Suppose $F(y|\mathbf{x})$ denotes the conditional distribution function of $Y$ at $\mathbf{X} = \mathbf{x}=(x_1,\ldots,x_d)$. Suppose $\{B_{j,m_1}(\cdot)\}_{j=1}^{p_1+m_1}$ denotes the B-spline coefficients of degree $m_1$ on the knot sequence $\{t_i\}_{i=0}^{p_1}$ as mentioned earlier. Then the conditional distribution $F(y|\mathbf{x})$ is given by
\begin{align}
F(y|\mathbf{x}) = \sum_{j=1}^{p_1+m_1}\phi_j(\mathbf{x})B_{j,m_1}(y)\; \text{where} \; 0=\phi_1(\mathbf{x})< \cdots < \phi_{p_1+m_1}(\mathbf{x})=1
\end{align}
It is noted that again the coefficients $\{\phi_j(\mathbf{x})\}_{j=1}^{p_1+m_1}$ are taken in such a way that the monotonicity of the distribution function is preserved. To put a prior on $\{\phi_j(\mathbf{x})\}_{j=1}^{p_1+m_1}$, we use the same technique as mentioned in Section \ref{npsqr_model}. Then $\phi_j(\mathbf{x})$ is given by
\begin{align}
\phi_j(\mathbf{x}) = \sum_{k_1=1}^{p_2+m_2} \cdots \sum_{k_d=1}^{p_2+m_2}\beta_{jk_1 \cdots k_d}B_{k_1,m_2}(x_1)\ldots B_{k_d,m_2}(x_d).
\end{align}
Hence the parameters to be estimated are given by 
\begin{align}
0=\beta_{1k_1\cdots k_d}<\cdots<\beta_{(p_1+m_1)k_1\cdots k_d}=1, \; (k_1,\ldots,k_d) \in \{1,\ldots,(p_2+m_2)\}^d.
\label{npdfsqr_parameter}
\end{align}.

\section{Likelihood Evaluation}
\label{likely_here}
In this section we describe the likelihood evaluation for both complete data and grouped data where only the frequencies of observations at each range of quantiles are given with the corresponding values of the explanatory variables.

\subsection{Complete Data}
\label{likely_here_complete}
Suppose that the explanatory and the response variables in the data are given by $\{Y_i\}_{i=1}^n$ and $\{\mathbf{X}_i\}_{i=1}^n = \{(X_{1i},\ldots,X_{di})\}_{i=1}^n$ where $n$ denotes the sample size and $d$ denotes the dimension of the explanatory variable. In the case of NPSQR, the likelihood derived from the quantile function is given by $\prod_{i=1}^{n} f(Y_i|\mathbf{X}_i)$ where $f(Y_i|\mathbf{X}_i)$ is given by
\begin{align*}
f(Y_i|\mathbf{X}_i) = \bigg(\frac{\partial}{\partial \tau}Q(\tau|\mathbf{X}_i)\bigg|_{\tau=\tau_{\mathbf{X}_i}(Y_i)}\bigg)^{-1}, \; i = 1,\ldots,n;
\end{align*}
here $\tau_{\mathbf{X}_i}(Y_i)$ solves the equation 
\begin{align}
Y_{i} = Q(\tau|\mathbf{X}_i)=\sum_{j=1}^{p_1+m_1}\theta_j(\mathbf{X}_i)B_{j,m_1}(\tau).
\label{eq:npsqr}
\end{align}
As described in Section \ref{npsqr_model}, $Q(\tau|\mathbf{X}_i)$ is constructed in such a way that it is monotonically increasing in $\tau$. Hence Equation \eqref{eq:npsqr} has a unique solution. In case we consider piece-wise quadratic B-spline (i.e., $m_1$=2), an advantage of our proposed method is that Equation \eqref{eq:npsqr} reduces to a quadratic equation and hence it can be solved analytically. Solving it analytically provides the exact solution in lesser time. Now to compute the likelihood, using the properties of derivative of B-spline (\cite{Boor2001}), we get
\begin{align}
\frac{\partial}{\partial t}Q(\tau|\mathbf{X}_i) = \frac{\partial}{\partial \tau}\sum_{j=1}^{p_1+m_1}\theta_j(\mathbf{X}_i)B_{j,m_1}(\tau) = \sum_{j=2}^{p_1+m_1}\theta^*_j(\mathbf{X}_i)B_{j-1,m_1-1}(\tau),
\label{eq:ddt_here}
\end{align}
where 
\begin{align*}
\theta^*_j(\mathbf{X}_i) = (p_1+m_1)(\theta_j(\mathbf{X}_i)-\theta_{j-1}(\mathbf{X}_i)),\;  j = 2,\ldots,p_1+m_1.
\end{align*}
Thus the log-likelihood in case of NPSQR is given by 
\begin{align}
\sum_{i=1}^{n}\log f(Y_i|\mathbf{X}_i) = & -\sum_{i=1}^{n}\log \bigg\{\sum_{j=2}^{p_1+m_1}\theta^*_j(\mathbf{X}_i)B_{j-1,m_1-1}(\tau_{\mathbf{X}_i}(Y_i)) \bigg\}.
\label{eq:likelihood_npsqr}
\end{align}
In case of NPDFSQR, the log-likelihood function is given by
\begin{align}
\sum_{i=1}^n \log f(Y_i|\mathbf{X}_i) = & \sum_{i=1}^n \log \bigg(\frac{\partial}{\partial y}F(y|\mathbf{X}_i)\bigg|_{y=Y_i}\bigg) \nonumber \\
= & \sum_{i=1}^n \log \bigg(\frac{\partial}{\partial y}\sum_{j=1}^{p_1+m_1}\phi_j(\mathbf{X}_i)B_{j,m_1}(y)\bigg|_{y=Y_i}\bigg) \nonumber \\
= & \sum_{i=1}^n \log \bigg(\sum_{j=2}^{p_1+m_1}\phi_j^*(\mathbf{X}_i)B_{j-1,m_1-1}(Y_i)\bigg),
\label{eq:likelihood_npdfsqr}
\end{align}
where 
\begin{align*}
\phi^*_j(\mathbf{X}_i) = (p_1+m_1)(\phi_j(\mathbf{X}_i)-\phi_{j-1}(\mathbf{X}_i)),\;  j = 2,\ldots,p_1+m_1.
\end{align*}

\subsection{Quantile Grid Data}
\label{likely_here_grid}
In the case of grid data suppose the partition of the quantiles of the response are given by $0=\rho_0<\rho_1<\cdots<\rho_c=1$ and for each observation it is given between which two consecutive quantile divisions it belongs. Define 
\begin{align}
I_{Y_i}(l) =\begin{cases}
1 \; & \text{if} \; Y_i \; \text{is in between} \; \rho_{l-1} \; \text{and} \; \rho_l\text{-th quantiles}\\
0 & \text{otherwise}
\end{cases}
\end{align}
for $i=1,\ldots,n$. Given the value of the explanatory variable for the $i$-th subject $\mathbf{X}_i$, the probability of an observation $Y_i$ belonging between $\rho_{l-1}$ and $\rho_l$-th quantile is given by $(F(q_Y(\rho_l)|\mathbf{X}_i) - F(q_Y(\rho_{l-1})|\mathbf{X}_i))$. Here $q_Y(g)$ denotes the $g$-th quantile $(0\leq g\leq 1)$ of $Y$. Hence the total likelihood is given by
\begin{align}
L = \prod_{i=1}^n\bigg(\prod_{l=1}^c(F(q_Y(\rho_l)|\mathbf{X}_i) - F(q_Y(\rho_{l-1})|\mathbf{X}_i))^{I_{Y_i}(l)}\bigg).
\label{eq:grid_likelihood}
\end{align}
We assume that the values $\{q_Y(\rho_l)\}_{l=1}^{c-1}$ are provided and with $F(q_Y(\rho_0)|\mathbf{X}_i) = 0$, $F(q_Y(\rho_c)|\mathbf{X}_i) = 1$ for all $i=1,\ldots,n$. Now for NPSQR, it should be noted that $F(q_Y(\rho_l)|\mathbf{X}_i)$ is the solution of the equation $Q(\tau|\mathbf{X}_i)$$= q_Y(\rho_l)$ since $Q(\tau|\mathbf{X}_i) = q_Y(\rho_l)$ implies $F(q_Y(\rho_l)|\mathbf{X}_i) = \tau$. Hence $F(q_Y(\rho_l)|\mathbf{X}_i)$ can be obtained solving the following equation in terms of $\tau$
\begin{align}
q_Y(\tau) = Q(\tau|\mathbf{X}_i)=\sum_{j=1}^{p_1+m_1}\theta_j(\mathbf{X}_i)B_{j,m_1}(\tau).
\label{eq:npsqr_grid}
\end{align}
In case of NPDFSQR, once the values of $\{Y_{\rho_l}\}_{l=0}^c$ are evaluated, likelihood evaluation is straightforward and can be easily obtained using Equation \eqref{eq:grid_likelihood}.

\section{Block Metropolis-Hastings MCMC Algorithm}
\label{MCMC_here}
To estimate the parameters for NPSQR and NPDFSQR methods (given by Equations \eqref{npsqr_parameter} and \eqref{npdfsqr_parameter} respectively) we use Block Metropolis-Hastings Markov Chain Monte Carlo algorithm (\cite{Chib1995}). It should be noted that the parameter space which needs to be estimated are of the same form for NPSQR and NPDFSQR and hence we use the similar steps and same prior distribution for both the cases. Recall that for NPSQR the parameter space is given by
$$0=\alpha_{1k_1\cdots k_d}<\cdots<\alpha_{(p_1+m_1)k_1\cdots k_d}=1, \; (k_1,\ldots,k_d) \in \{1,\ldots,(p_2+m_2)\}^d.$$ Now define $$ \gamma_{jk_1\cdots k_d} = \alpha_{(j+1)k_1\cdots k_d} - \alpha_{jk_1\cdots k_d},\; j=1,\ldots,p_1+m_1-1$$ for $\{k_1,\ldots, k_d\} \in \{1,\ldots, (p_2+m_2)\}^d$. Then it can be noted that $\{\gamma_{jk_1\cdots k_d}\}_{j=1}^{p_1+m_1-1}$ belongs to the unit-simplex for any given $\{k_1,\ldots, k_d\} \in \{1,\ldots, (p_2+m_2)\}^d$ since $$\sum_{j=1}^{p_1+m_1}\gamma_{jk_1\cdots k_d} = 1,\; \gamma_{jk_1\cdots k_d} \geq 0, \; j= 1,\ldots, p_1+m_1-1,$$for $\{k_1,\ldots, k_d\} \in \{1,\ldots, (p_2+m_2)\}^d$. We put the uniform prior on each simplex block. The number of simplex blocks is given by $(p_2+m_2)^d$ where $\{s_i\}_{i=0}^{p_2}$ is the equidistant knot sequence used for each coordinate of the explanatory variables and $m_2$ is the degree of the piece-wise polynomials use for B-spline regression (as mentioned in Section \ref{npsqr_model}). Within an iteration, one simplex block is updated at a time. Hence there will be $(p_2+m_2)^d$ updates performed one at a time during a single iteration.

To make movements on the simplex block, we use the same strategy as explained in \cite{DasQR_1_2016} and \cite{DasQR_2_2016}. First we fix $\{k_1,\ldots, k_d\} \in \{1,\ldots, (p_2+m_2)\}^d$. We generate independent sequence $\{U_j\}_{j=1}^{p_1+m_1-1}$ from $U(1/r,r)$ for some $r>1$. It should be noted that $r$ works as a tuning parameter of the MCMC. Smaller value of $r$ yields sticky movement with higher acceptance probability while larger value of $r$ would result in bigger jumps with less acceptance probability. Define $V_j=\gamma_{jk_1\cdots k_d} U_j$. Hence the proposal move $\gamma_{jk_1\cdots k_d} \mapsto \gamma_{jk_1\cdots k_d}^*$ is given by
\begin{align*}
\gamma_{jk_1\cdots k_d}^* = \frac{V_j}{\sum_{i=1}^{p_1+m_1-1}V_i},\; j=1,\ldots,p_1+m_1-1. 
\end{align*}
The conditional distribution of $\{\gamma_{jk_1\cdots k_d}^*\}_{j=1}^{p_1+m_1-1}$ given $\{\gamma_{jk_1\cdots k_d}\}_{j=1}^{p_1+m_1-1}$ is given by (see Section \ref{section_mcmc_1} for the derivation)
\begin{align}
f(\mathbf{\gamma}_{.k_1\cdots k_d}^*|\mathbf{\gamma}_{.k_1\cdots k_d})= \bigg(\frac{r}{r^2-1}\bigg)^{p_1+m_1-1}\bigg\{\prod\limits_{j=1}^{p_1+m_1-1} \gamma_{jk_1\cdots k_d}\bigg\}^{-1}\frac{(D_1-D_2)}{(p_1+m_1-1)},
\end{align}
where 
\begin{align*}
D_1 = & \Big(\min_{0 \leq j \leq p_1+m_1-1}\ \frac{r\gamma_{jk_1\cdots k_d}}{\gamma_{jk_1\cdots k_d}^*}\Big)^{p_1+m_1-1}, \\ 
D_2 = & \Big(\max_{0 \leq j \leq p_1+m_1-1} \frac{\gamma_{jk_1\cdots k_d}}{r\gamma_{jk_1\cdots k_d}^*}\Big)^{p_1+m_1-1}.
\end{align*}
Suppose $L(\gamma_{.k_1\cdots k_d})$ and $L(\gamma^*_{.k_1\cdots k_d})$ denote the likelihood for NPSQR (either corresponding to complete data or quantile grid data) for the parameter values $\gamma_{.k_1\cdots k_d} = \{\gamma_{jk_1\cdots k_d}\}_{j=1}^{p_1+m_1-1}$ and $\gamma^*_{.k_1\cdots k_d} = \{\gamma_{jk_1\cdots k_d}\}_{j=1}^{p_1+m_1-1}$ respectively. Then a single block update for $(k_1,\ldots, k_d) \in \{1,\ldots, (p_2+m_2)\}^d$ is given by $P_{k_1\cdots k_d}= \min \{p_{k_1\cdots k_d}, 1\}$ where 
\begin{align*}
p_{k_1\cdots k_d} = &\frac{L(\gamma^*_{.k_1\cdots k_d})\pi(\gamma^*_{.k_1\cdots k_d})f(\mathbf{\gamma}_{.k_1\cdots k_d}|\mathbf{\gamma}^*_{.k_1\cdots k_d})}{L(\gamma_{.k_1\cdots k_d})\pi(\gamma_{.k_1\cdots k_d})f(\mathbf{\gamma}_{.k_1\cdots k_d}^*|\mathbf{\gamma}_{.k_1\cdots k_d})},\\
= &\frac{L(\gamma^*_{.k_1\cdots k_d})f(\mathbf{\gamma}_{.k_1\cdots k_d}|\mathbf{\gamma}^*_{.k_1\cdots k_d})}{L(\gamma_{.k_1\cdots k_d})f(\mathbf{\gamma}_{.k_1\cdots k_d}^*|\mathbf{\gamma}_{.k_1\cdots k_d})},\\
\end{align*}
and $\pi(\cdot)$ denotes the uniform prior density.

For NPDFSQR we define $$ \delta_{jk_1\cdots k_d} = \beta_{(j+1)k_1\cdots k_d} - \beta_{jk_1\cdots k_d},\; j=1,\ldots,p_1+m_1-1$$ for $\{k_1,\ldots, k_d\} \in \{1,\ldots, (p_2+m_2)\}^d$. Hence in this case also the parameter space is given by a collection of $(p_2+m_2)^d$ simplex blocks. Hence in case of NPDFSQR (for complete data and quantile grid data), the update steps are performed similar to that of NPSQR (as mentioned above). In this case also, inside each iteration step of MCMC, $(p_2+m_2)^d$ simplex blocks are updated one by one.
\subsection{Warm Start}
\label{warm_start}
 In case of very large parameter space, the strategy of warm-start in general helps in reducing the burn-in for Metropolis-Hastings MCMC. In the proposed method, instead of using a randomly generated starting point, we use the maximum likelihood estimator (MLE) as the starting point. In Section \ref{MCMC_here}, it is noted that for both NPSQR and NPDFSQR methods, the parameter space is given by a collection of simplex blocks. One of the challenging aspect of finding MLE under this scenario is that the parameter space is constrained and for NPSQR, the likelihood function does not have any closed form. It is almost impossible to verify whether the negative of the likelihood of NPSQR is convex or not (in case negative likelihood is convex, there exists only one global maximum and convex optimization techniques can be used to find it). Thus optimization should be performed assuming the possibility of existence of multiple local maximums of the likelihood function. Secondly, due to the absence of a closed form likelihood in the case of NPSQR, the derivative of the likelihood function does not have any closed form. In this case, one of the disadvantages of using derivative based optimization methods is that derivatives can be evaluated only numerically which is computationally intensive. Hence, this is an ideal scenario to use black-box optimization technique since black-box optimization technique does not use analytically derivative and is used to optimize any function with (possibly) multiple maximums or minimums.
 
 Recently \cite{DAS_1_2016} proposed an black-box optimization technique on a hyper-rectangular parameter space which has been shown to perform better (in terms of computing time, accuracy and successful convergence) than black-box optimization techniques Genetic Algorithm (\cite{Fraser1957}, \cite{Bethke1980}, \cite{Goldberg1989}) \hspace{.2cm} and \hspace{.2cm} Simulated \hspace{.2cm} Annealing \hspace{.2cm} (\cite{Kirkpatrick1983}, \cite{Granville1994}) yielding better solutions. \hspace{.2cm} Following \hspace{.2cm} that \hspace{.2cm} strategy, \cite{DAS_2_2016} modified that algorithm to optimize any function on an unit simplex. Later \cite{DAS_3_2016} extended that method and proposed `Greedy Coordinate Descent of Varying Step sizes on Multiple Simplexes' (GCDVSMS) algorithm which efficiently minimizes (or maximizes) any black-box function of parameters given by a collection of unit simplex blocks. The main idea of the this algorithm is to make jumps of varying step-sizes within each unit simplex blocks parallelly and searching for the most favorable direction of movement. We use GCDVSMS algorithm to find the warm starting point before initializing the MCMC.

\subsection{Automatic Controlling of Acceptance Probability}
\label{acc_prob_here}
As mentioned Section \ref{MCMC_here}, $r$ plays a critical role in controlling the acceptance probability. Instead of fixing the value of $r$, we propose an adaptive strategy so that during the MCMC iterations, acceptance probability is maintained within a desirable range of 0.15 to 0.45.

For both NPSQR and NPDFSQR, we start the first iteration with $r=1.05$. It should be noted that in each iteration, $(p_2+m_2)^d$ simplex blocks are updated; hence, $(p_2+m_2)^d$ acceptance-rejection decisions are taken. After each iteration, the cumulative acceptance ratio is calculated. At the end of an iteration, if the cumulative acceptance probability drops below 0.15, $r$ is updated to $1+(r-1)/2$ and if the cumulative acceptance probability goes above 0.45, the value of $r$ is updated to $1+2(r-1)$. Note that in this way, the value of $r$ will always be greater than 1.

\section{Simulation study}
\label{sim_study}
For simulation purpose, we consider the following true models and we generate sample of sizes $n=50,100,200$ for each case.
\begin{enumerate}[label=(\Alph*)]
\item \textbf{First Simulation Study :} We consider the explanatory variable $X$ in coming from $U(0,5)$. The dependence of the response variable $Y$ on $X$ is given by
\begin{align}
y_i = x_i + \sin (2x_i) + 3\epsilon_i, \; i=1,\ldots,n.
\label{sim_1_eq}
\end{align}
where $\epsilon_i$ follows $SN(4; 0, 1)$ (i.e., skewed normal distribution with scale parameter 4, mean 0 and standard deviation 1). 
\item \textbf{Second Simulation Study :} We consider the explanatory variable $X$ in coming from $U(-100,100)$. The dependence of the response variable $Y$ on $X$ is given by
\begin{align}
y_i = -x_i^3/100000 + (\sin (\pi x_i/100)+4)U_i\epsilon_i \; \text{for} \; i=1,\ldots,n.
\label{sim_2_eq}
\end{align}
where $U_i$ follows discrete uniform $\{-1,1\}$ and $\epsilon_i$ follows gamma distribution with shape and scale parameters 5 and 1 respectively.
\end{enumerate}

\subsection{Case of Complete Data}
\label{sec_complete}
For each of the cases given by Equation (\ref{sim_1_eq}) and (\ref{sim_2_eq}), the quantiles are estimated using Non-parametric Simultaneous Quantile Regression (NPSQR), Non-parametric Distribution Function Simultaneous Quantile Regression (NPDFSQR), Local Linear Quantile Regression (LLQR)  and  Local Spline Quantile Regression (LSQR) (see \cite{Yu1998}, \cite{Koenker2015}). \\

For NPSQR and NPDFSQR, first the explanatory variable and the response variables are transformed into unit interval separately by linear transformation. We use piece-wise quadratic B-spline for expanding both the quantile function as well as the explanatory variable (i.e., $m_1=m_2=2$). One of the biggest advantage of quadratic B-spline is that Equations \eqref{eq:npsqr} and \eqref{eq:npsqr_grid} reduce to quadratic equation which can be solved analytically very easily. Suppose $\{t_i\}_{i=0}^{p_1}$ and $\{s_i\}_{i=0}^{p_2}$ denote the equidistant knots on unit interval such that
\begin{align*}
0=t_0<t_1<\cdots<t_{p_1}=1, (t_i-t_{i-1}) = \frac{1}{p_1} \; \text{for} \; i=1,\ldots,p_1, \\
0=s_0<s_1<\cdots<s_{p_2}=1, (s_i-s_{i-1}) = \frac{1}{p_2} \; \text{for} \; i=1,\ldots,p_2.
\end{align*}
In case of NPSQR, for B-spline expansion, we consider $\{t_i\}_{i=0}^{p_1}$ and $\{s_i\}_{i=0}^{p_2}$ knot sequences for expanding the quantile function and transformed explanatory variable respectively. We consider $p_1=p_2=3,\ldots, 10$ and choose the best model using the Akaike Information Criteria (AIC) criterion. For NPDFSQR, suppose that knot sequences $\{t_i\}_{i=0}^{p_1}$ and $\{s_i\}_{i=0}^{p_2}$ are used for expanding the distribution function and transformed explanatory variable respectively. For NPDFSQR, it is noted that taking $p_1=p_2=3,4$, yields poorly estimated posterior quantile curves under different simulation studies. The possible reason is that in case of NPDFSQR, a lot of estimated B-spline coefficients (which are used to estimate the distribution function of the response variable) are coming to be nearly zero which might be an indicator of usage of less number of knots to represent the variability of the distribution function. Therefore, for NPDFSQR, we omit the cases $p_1=p_2=3,4$ and consider the cases $p_1=p_2=5,\ldots, 10$. Then we choose the best model via the AIC criterion.

\begin{figure}[] 
   \begin{subfigure}[b]{0.5\linewidth}
    \centering
    \includegraphics[width=0.99\linewidth]{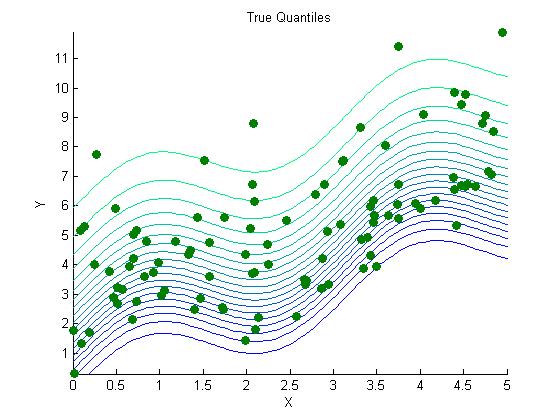} 
    \caption{True Quantiles}  
    \label{1_complete_TRUE} 
  \end{subfigure} 
   \begin{subfigure}[b]{0.5\linewidth}
    \centering
    \includegraphics[width=0.20\linewidth]{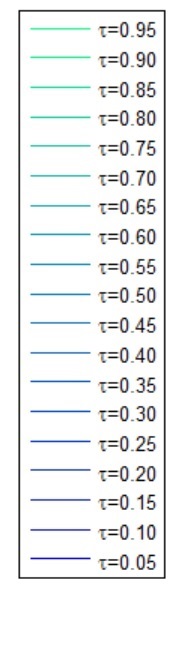} 
   \caption{Quantile legend}  
    \label{1_complete_QUANTILES} 
  \end{subfigure}
\begin{subfigure}[b]{0.5\linewidth}
    \centering
    \includegraphics[width=0.99\linewidth]{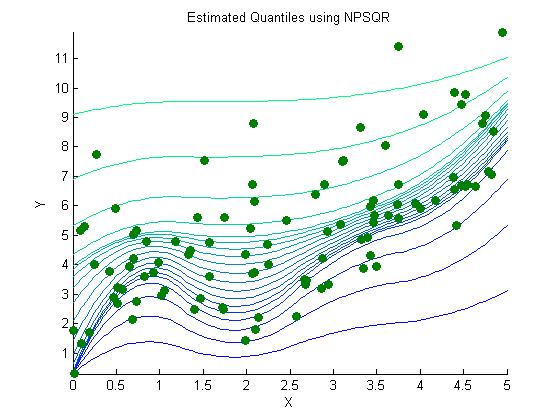} 
    \caption{NPSQR Estimated Quantiles} 
    \label{1_complete_NPSQR} 
  \end{subfigure}
  \begin{subfigure}[b]{0.5\linewidth}
    \centering
    \includegraphics[width=0.99\linewidth]{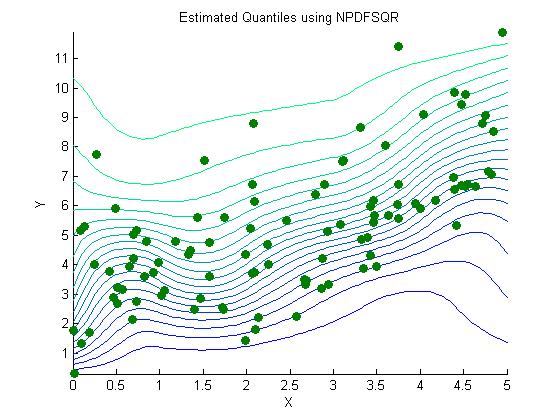} 
    \caption{NPDFSQR Estimated Quantiles} 
    \label{1_complete_NPDFSQR} 
  \end{subfigure} \\
  \begin{subfigure}[b]{0.5\linewidth}
    \centering
    \includegraphics[width=0.99\linewidth]{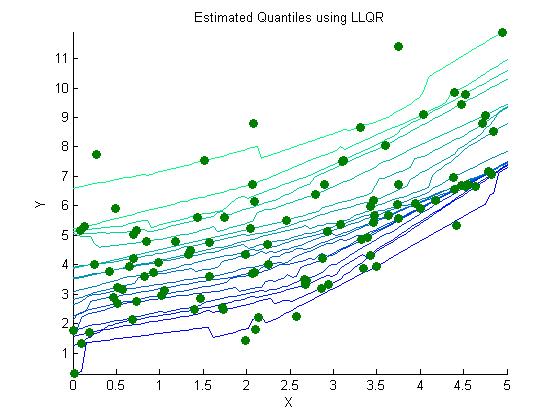} 
    \caption{LLQR Estimated Quantiles} 
    \label{1_complete_LLQR} 
  \end{subfigure}
    \begin{subfigure}[b]{0.5\linewidth}
    \centering
    \includegraphics[width=0.99\linewidth]{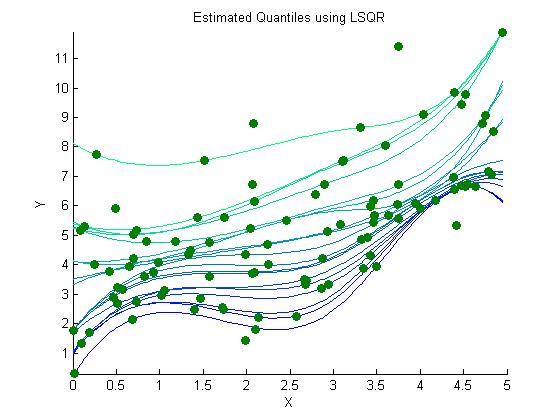} 
    \caption{LSQR Estimated Quantiles}  
    \label{1_complete_LSQR} 
  \end{subfigure} \\
  \caption{(First Simulation Study (Complete Data)) True and estimated quantiles at $\tau = \{0.05, 0.10, 0.15, \ldots, 0.90, 0.95\}$ for $n=100$ using NPSQR, NPDFSQR, LLQR and LSQR with the data points.}
  \label{1_COMPLETE} 
\end{figure}

\begin{table}[]
\centering
\begin{tabular}{@{}ccccc@{}}
\toprule
\begin{tabular}[c]{@{}c@{}}Sample\\ size\end{tabular} & NPSQR & NPDFSQR & LLQR & LSQR \\ \midrule
$n=50$ & 4.90 & 4.66 & 5.13 & 5.42 \\
$n=100$ & 4.40 & 4.51 & 4.59 & 4.50 \\
$n=200$ & 4.34 & 4.47 & 4.65 & 4.56 \\ \bottomrule
\end{tabular}
\caption{(First Simulation Study (Complete Data)) Prediction Mean Squared Errors using NPSQR, NPDFSQR, LLQR and LSQR based on simulation studies for sample sizes $n=50, 100, 200$.}
\label{tab_complete_1}
\end{table}

As mentioned in Section \ref{warm_start}, for all considered MCMC schemes in this paper, we use GCDVSMS algorithm to find the starting point. The values of the tuning parameters in the GCDVSMS algorithm have been taken to be as follows : \textit{initial global step size} $s_{initial} = 1$, \textit{step decay rate} for the first \textit{run} $\rho_1=2$, \textit{step decay rate} for other \textit{runs} $\rho_2 = 1.05$, \textit{step size threshold} $\phi = 10^{-2}$, \textit{sparsity threshold} $\lambda=10^{-3}$, the convergence criteria controlling parameters \textit{tol\_fun\_1}=\textit{tol\_fun\_2}$=10^{-2}$,  maximum number of iterations inside each \textit{run} $\textit{max\_iter} = 5000$, maximum number of allowed \textit{runs} $\textit{max\_runs}=200$. For both NPSQR and NPDFSQR, we perform 10000 iterations discarding first 1000 iterations as burn-in. After the quantile curves are estimated, inverse linear transformations are applied on the response and the explanatory variables to get them back to their original scales.

\begin{figure}[] 
   \begin{subfigure}[b]{0.5\linewidth}
    \centering
    \includegraphics[width=0.99\linewidth]{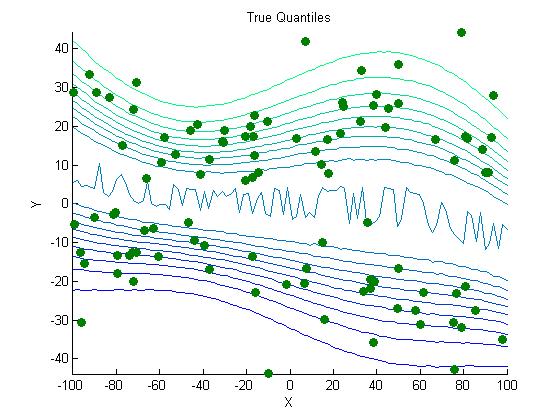} 
    \caption{True Quantiles}  
    \label{2_complete_TRUE} 
  \end{subfigure} 
   \begin{subfigure}[b]{0.5\linewidth}
    \centering
    \includegraphics[width=0.20\linewidth]{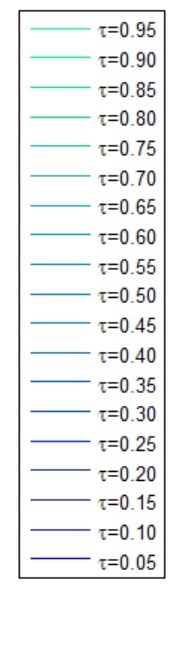} 
   \caption{Quantile legend}  
    \label{2_complete_QUANTILES} 
  \end{subfigure}
\begin{subfigure}[b]{0.5\linewidth}
    \centering
    \includegraphics[width=0.99\linewidth]{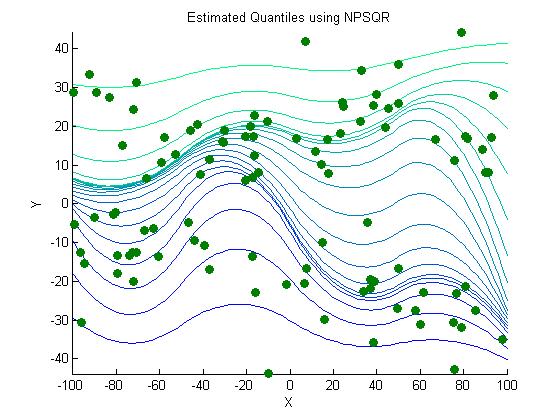} 
    \caption{NPSQR Estimated Quantiles} 
    \label{2_complete_NPSQR} 
  \end{subfigure}
  \begin{subfigure}[b]{0.5\linewidth}
    \centering
    \includegraphics[width=0.99\linewidth]{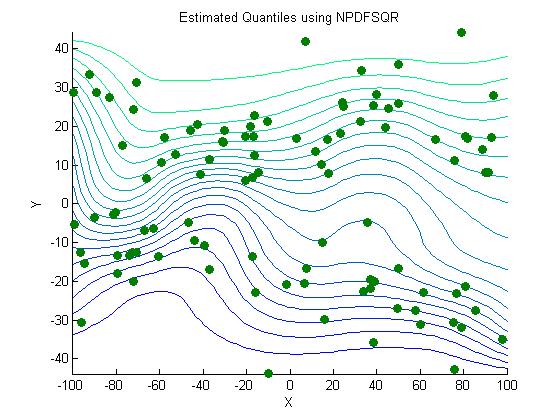} 
    \caption{NPDFSQR Estimated Quantiles} 
    \label{2_complete_NPDFSQR} 
  \end{subfigure} \\
  \begin{subfigure}[b]{0.5\linewidth}
    \centering
    \includegraphics[width=0.99\linewidth]{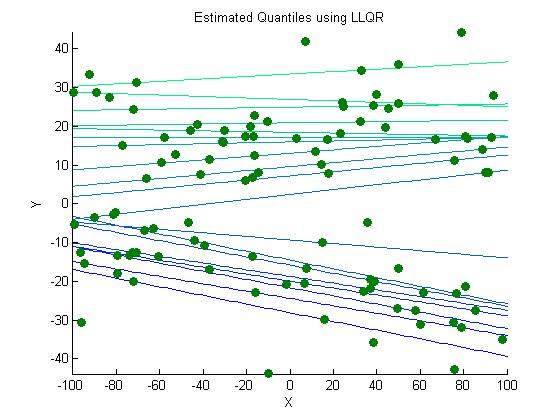} 
    \caption{LLQR Estimated Quantiles} 
    \label{2_complete_LLQR} 
  \end{subfigure}
    \begin{subfigure}[b]{0.5\linewidth}
    \centering
    \includegraphics[width=0.99\linewidth]{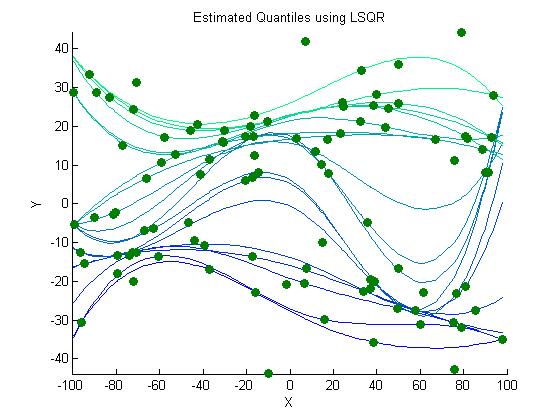} 
    \caption{LSQR Estimated Quantiles}  
    \label{2_complete_LSQR} 
  \end{subfigure} \\
  \caption{(Second Simulation Study (Complete Data)) True and estimated quantiles at $\tau = \{0.05, 0.10, 0.15, \ldots, 0.90, 0.95\}$ for $n=100$ using NPSQR, NPDFSQR, LLQR and LSQR with the data points.}
  \label{2_COMPLETE} 
\end{figure}

\begin{table}[]
\centering
\begin{tabular}{@{}ccccc@{}}
\toprule
\begin{tabular}[c]{@{}c@{}}Sample\\ size\end{tabular} & NPSQR & NPDFSQR & LLQR & LSQR \\ \midrule
$n=50$ & 564.81 & 576.18 & 576.84 & 641.17 \\
$n=100$ & 559.14 & 539.68 & 592.83 & 645.58 \\
$n=200$ & 504.22 & 517.63 & 532.97 & 554.40 \\ \bottomrule
\end{tabular}
\caption{(Second Simulation Study (Complete Data)) Prediction Mean Squared Errors using NPSQR, NPDFSQR, LLQR and LSQR based on simulation studies for sample sizes $n=50, 100, 200$.}
\label{tab_complete_2}
\end{table}

In NPDFSQR, once the whole distribution function is estimated non- parametrically, the quantile function is obtained evaluating numerically. We take a grid of length 1000 on transformed $Y$ variable which is a unit interval. For any given value of $X=x$, the distribution function is evaluated at these 1000 equidistant grid-points. Then $Q(\tau|x)$ is estimated using interpolation from the values of the distribution function at those aforementioned 1000 points.

For any given quantile level, LSQR (\cite{Koenker2015}) fits a piecewise cubic polynomial with any given number of knots (breakpoints in the third derivative) arranged at that quantile of the $X$. However, no explicit way for deciding the number of knots for a given data set has been provided in that article. For fair comparison, while fitting LSQR, we consider the same number of knots as used in estimating the quantile levels with NPSQR for that data-set. (Note that, as mentioned earlier, the number of knots in NPSQR is selected based on the AIC criterion). While estimating any given quantile level with LLQR, to select the bandwidth, we follow the technique mentioned in the section 2 of \cite{Yu1998}. We use `\texttt{quantreg}' (\cite{quantreg}) R-package for LLQR and LSQR. Except for the bandwidth selection for LLQR, rest of the codes have been followed as provided in \cite{Koenker2015}.

For comparing the performances of the proposed methods, LLQR and LSQR, 1000 pairs of observations $\{(X_i,Y_i)\}_{i=1}^{1000}$ are generated from Equation \eqref{sim_1_eq}. Let $\hat{Q}(\tau|x)$ denote the estimated value of the $\tau$-th quantile at $X=x$. Then the Prediction Mean Squared Error (PMSE) is given by
$$\text{PMSE} = \frac{1}{1000}\sum_{i=1}^{1000} (Y_i - \hat{Q}(0.5|X_i) )^2.$$
Note that $\hat{Q}(0.5|X_i)$ denotes the estimated median estimate at $X=X_i$. Except for LSQR, it is straightforward to find $\hat{Q}(0.5|X_i)$ for any given $X_i$. The way LSQR is performed in \cite{Koenker2015}, the quantile curves are evaluated only at those points where $X$ is given (in the data). So in this case, we use linear interpolation with  \texttt{interp} function in R to find the approximate values of $\{\hat{Q}(0.5|X_i)\}_{i=1}^{1000}$ from the estimated median values at the points given in the data.

In the simulation study, in Tables \ref{tab_complete_1} and \ref{tab_complete_2}, it is noted that the performances of NPSQR and NPDFSQR are generally better than LLQR and LSQR in terms of PMSE. For both NPSQR and NPDFSQR, we note an overall decreasing trend of PMSE with increasing sample size. We note that unlike the case of the proposed methods, using LLQR and LSQR the estimated quantile lines cross each other. It is also noted that the estimated quantile curves using LSQR in the simulation studies (and also in the example provided in \cite{Koenker2015}) have a tendency to pass through the data points which may not be desirable specially for estimating quantile curves with small sample.

\subsection{Case of Grid data}
\label{sec_grid}
In case of grid data, we obtain the grid data by coarsening the data generated in Section \ref{sec_complete} into grid data. While transforming a sample of given size into grid data, we consider three types of grid data generated from each sample which are 5, 10 and 20 percentile gap grid data. For example, in case of 5 percentile gird data, the values of $\{q_Y(\rho_l)\}_{l=1}^{19}$ are given where $\rho_l=0.05*l$ for $l=1,\ldots,19$. The values of $\{q_Y(\rho_l)\}_{l=1}^{19}$ are computed non-parametrically from the given sample using \texttt{quantile} function in MATLAB (version R2014a). Then for each given value of $X_i$, it is noted that between which two consecutive quantile girds the corresponding observation $Y_i$ belongs for $i = 1,\ldots,n$. Thus corresponding to each value of $X_i$, the position of $Y_i$ with respect to the quantile grids is given in the grid data. 10 and 20 percentile grid data are also generated in the similar way.

After the grid data is generated, we transform the values of explanatory and the response values into unit intervals separately using linear transformations. Once they are transformed into the unit intervals, the likelihood can be computed as described in Section \ref{likely_here_grid}. As mentioned earlier, except the likelihood evaluation part, the remaining part of the Block Metropolis-Hastings MCMC algorithm is similar to that of the case of complete data. Before starting the MCMC, we compute a warm starting point using GCDVSMS algorithm (\cite{DAS_3_2016}) with the values of the tuning parameters as mentioned in Section \ref{sec_complete}. We estimate the quantile curves for each cases using NPSQR and NPDFSQR methods. We also compute the PMSE (as defined in Section \ref{sec_complete}) for comparison under each scenario. For each cases, we perform 10000 MCMC iterations discarding the first 1000 iterations as burn-in. Similar to the case of complete data, for NPSQR we consider $p_1=p_2=3,\ldots, 10$ and for NPDFSQR we consider $p_1=p_2=5,\ldots, 10$. The best possible value of $p_1$ and $p_2$ in either cases are selected based on AIC criteria. After the quantile curves are estimated, inverse transformations on the response and explanatory variables are performed to return back to the original scale.

We note that NPSQR performs slightly better than NPDFSQR in terms of PMSE for both the simulation studies considered. It is also noted that with increasing sample size, there is a decreasing trend of PMSE for both the cases. Also, with smaller percentile gap data, PMSE comes out to be smaller in most of the cases for both NPSQR and NPDFSQR. To study the relative performance of estimating the quantile curves with proposed methods for complete and grid data, readers can compare the PMSE values in Tables \ref{tab_complete_1} and  \ref{tab_grid_1} (for the first simulation study) and in Tables \ref{tab_complete_2} and  \ref{tab_grid_2} (for second simulation study).
\begin{figure}[] 
\centering
   \begin{subfigure}[b]{0.4\linewidth}
    \centering
    \includegraphics[width=0.99\linewidth]{1_complete_TRUE.jpg} 
    \caption{True Quantiles}  
    \label{1_GRID_TRUE} 
  \end{subfigure} 
   \begin{subfigure}[b]{0.4\linewidth}
    \centering
    \includegraphics[width=0.20\linewidth]{1_complete_QUANTILES.jpg} 
   \caption{Quantile legend}  
    \label{1_GRID_QUANTILES} 
  \end{subfigure}
\begin{subfigure}[b]{0.4\linewidth}
    \centering
    \includegraphics[width=0.99\linewidth]{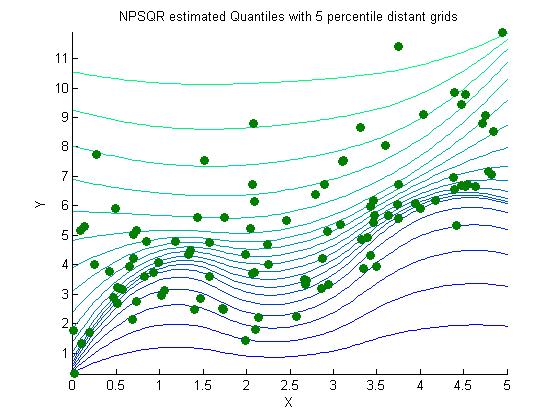} 
    \caption{NPSQR (5 percentile grid data)} 
    \label{1_truncated_NPSQR_5} 
  \end{subfigure}
  \begin{subfigure}[b]{0.4\linewidth}
    \centering
    \includegraphics[width=0.99\linewidth]{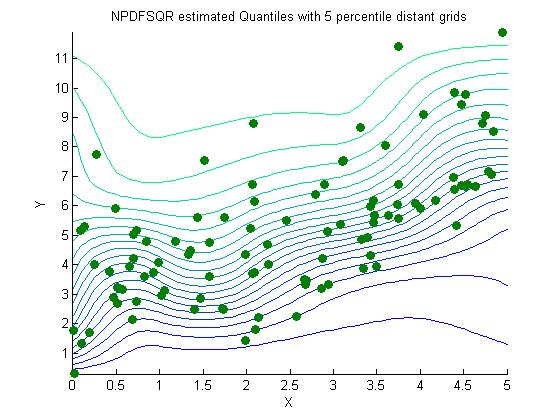} 
    \caption{NPDFSQR (5 percentile grid data)} 
    \label{1_truncated_NPDFSQR_5} 
  \end{subfigure} \\
  \begin{subfigure}[b]{0.4\linewidth}
    \centering
    \includegraphics[width=0.99\linewidth]{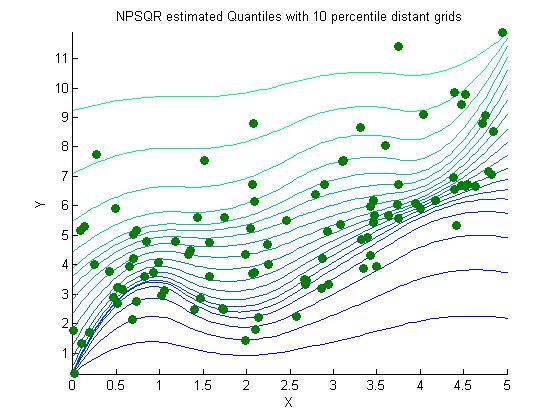} 
    \caption{NPSQR (10 percentile grid data)} 
    \label{1_truncated_NPSQR_10} 
  \end{subfigure}
    \begin{subfigure}[b]{0.4\linewidth}
    \centering
    \includegraphics[width=0.99\linewidth]{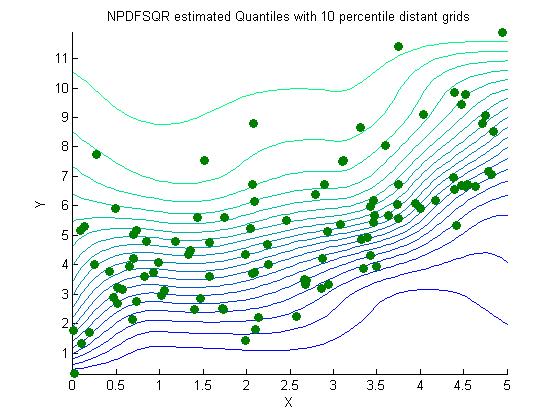} 
    \caption{NPDFSQR (10 percentile grid data)}  
    \label{1_truncated_NPDFSQR_10} 
  \end{subfigure} \\
    \begin{subfigure}[b]{0.4\linewidth}
    \centering
    \includegraphics[width=0.99\linewidth]{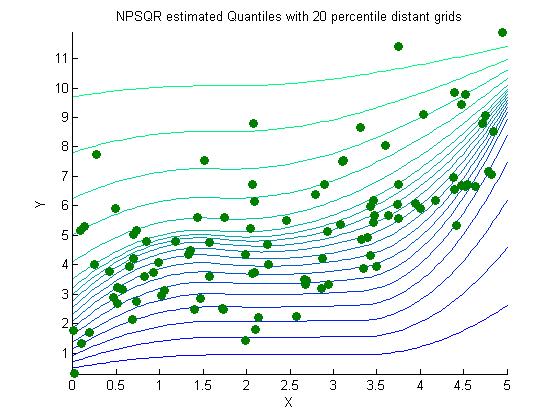} 
    \caption{NPSQR (20 percentile grid data)} 
    \label{1_truncated_NPSQR_20} 
  \end{subfigure}
    \begin{subfigure}[b]{0.4\linewidth}
    \centering
    \includegraphics[width=0.99\linewidth]{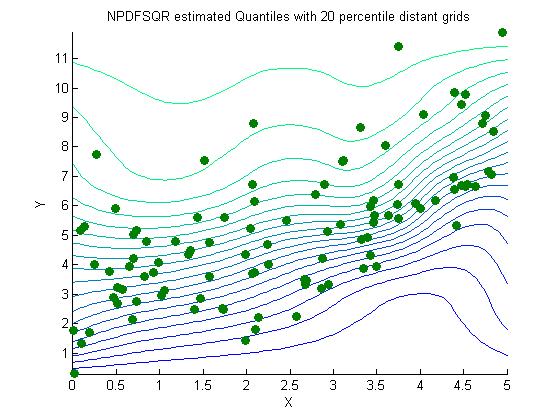} 
    \caption{NPDFSQR (20 percentile grid data)}  
    \label{1_truncated_NPDFSQR_20} 
  \end{subfigure} \\
  \caption{(First Simulation Study (Grid Data)) True and estimated quantiles at $\tau = \{0.05, 0.10, 0.15, \ldots, 0.90, 0.95\}$ for $n=100$ using NPSQR and NPDFSQR with the data points.}
  \label{1_GRID} 
\end{figure}

\begin{table}[]
\centering
\begin{tabular}{@{}cccc@{}}
\toprule
Sample size & \begin{tabular}[c]{@{}c@{}}Percentile gap\\ for grid data\end{tabular} & \begin{tabular}[c]{@{}c@{}}NPSQR\\ (pred error)\end{tabular} & \begin{tabular}[c]{@{}c@{}}NPDFSQR\\ (pred error)\end{tabular} \\ \midrule
\multirow{3}{*}{$n=50$} & 5 & 4.57 & 4.81 \\ \cmidrule(l){2-4} 
 & 10 & 4.64 & 4.75 \\ \cmidrule(l){2-4} 
 & 20 & 4.66 & 4.78 \\ \midrule
\multirow{3}{*}{$n=100$} & 5 & 4.30 & 4.55 \\ \cmidrule(l){2-4} 
 & 10 & 4.39 & 4.58 \\ \cmidrule(l){2-4} 
 & 20 & 4.77 & 4.70 \\ \midrule
\multirow{3}{*}{$n=200$} & 5 & 4.34 & 4.45 \\ \cmidrule(l){2-4} 
 & 10 & 4.40 & 4.47 \\ \cmidrule(l){2-4} 
 & 20 & 4.33 & 4.72 \\ \bottomrule
\end{tabular}
\caption{(First Simulation Study (Grid Data)) Prediction Mean Squared Errors using NPSQR and NPDFSQR based on simulation studies with $5,10,20$ distant percentile grids for sample sizes $n=50, 100, 200$.}
\label{tab_grid_1}
\end{table}

\begin{figure}[] 
\centering
   \begin{subfigure}[b]{0.4\linewidth}
    \centering
    \includegraphics[width=0.99\linewidth]{2_complete_TRUE.jpg} 
    \caption{True Quantiles}  
    \label{2_GRID_TRUE} 
  \end{subfigure} 
   \begin{subfigure}[b]{0.4\linewidth}
    \centering
    \includegraphics[width=0.20\linewidth]{1_complete_QUANTILES.jpg} 
   \caption{Quantile legend}  
    \label{2_GRID_QUANTILES} 
  \end{subfigure}
\begin{subfigure}[b]{0.4\linewidth}
    \centering
    \includegraphics[width=0.99\linewidth]{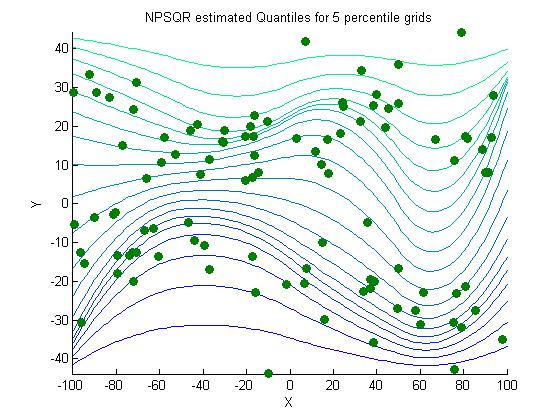} 
    \caption{NPSQR (5 percentile grid data)} 
    \label{2_truncated_NPSQR_5} 
  \end{subfigure}
  \begin{subfigure}[b]{0.4\linewidth}
    \centering
    \includegraphics[width=0.99\linewidth]{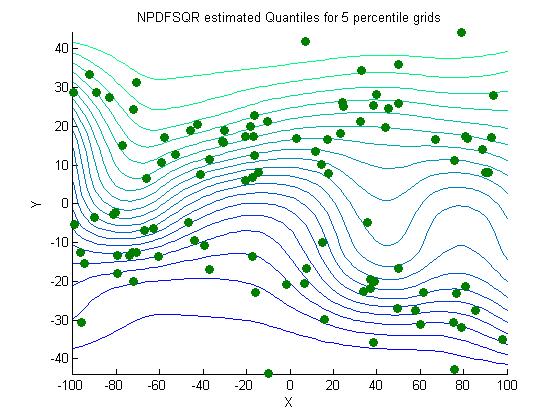} 
    \caption{NPDFSQR (5 percentile grid data)} 
    \label{2_truncated_NPDFSQR_5} 
  \end{subfigure} \\
  \begin{subfigure}[b]{0.4\linewidth}
    \centering
    \includegraphics[width=0.99\linewidth]{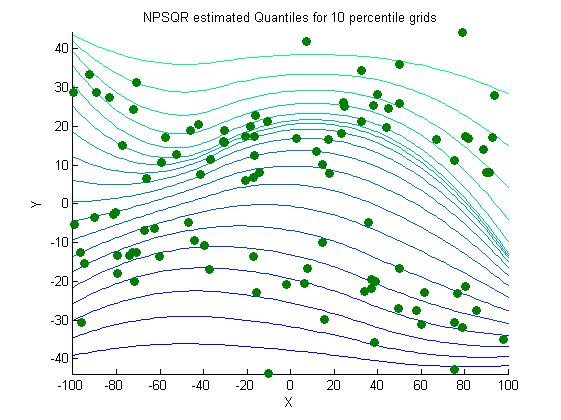} 
    \caption{NPSQR (10 percentile grid data)} 
    \label{2_truncated_NPSQR_10} 
  \end{subfigure}
    \begin{subfigure}[b]{0.4\linewidth}
    \centering
    \includegraphics[width=0.99\linewidth]{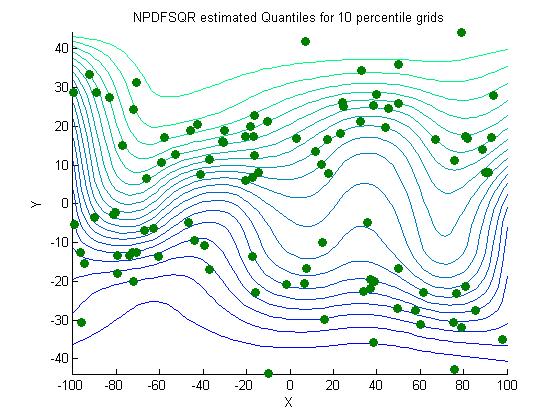} 
    \caption{NPDFSQR (10 percentile grid data)}  
    \label{2_truncated_NPDFSQR_10} 
  \end{subfigure} \\
    \begin{subfigure}[b]{0.4\linewidth}
    \centering
    \includegraphics[width=0.99\linewidth]{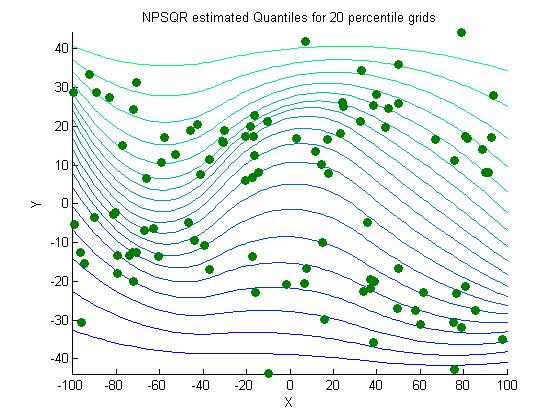} 
    \caption{NPSQR (20 percentile grid data)} 
    \label{2_truncated_NPSQR_20} 
  \end{subfigure}
    \begin{subfigure}[b]{0.4\linewidth}
    \centering
    \includegraphics[width=0.99\linewidth]{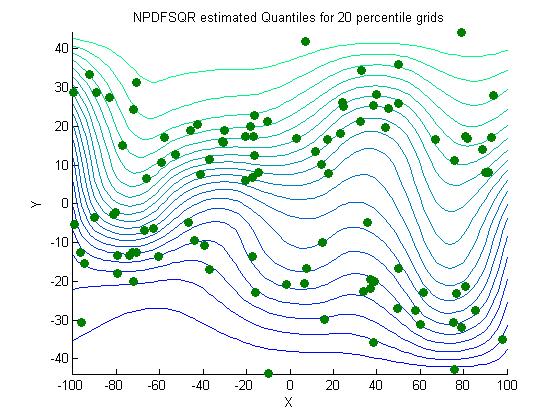} 
    \caption{NPDFSQR (20 percentile grid data)}  
    \label{2_truncated_NPDFSQR_20} 
  \end{subfigure} \\
  \caption{(Second Simulation Study (Grid Data)) True and estimated quantiles at $\tau = \{0.05, 0.10, 0.15, \ldots, 0.90, 0.95\}$ for $n=100$ using NPSQR and NPDFSQR with the data points.}
  \label{2_GRID} 
\end{figure}

\begin{table}[]
\centering
\begin{tabular}{@{}cccc@{}}
\toprule
Sample size & \begin{tabular}[c]{@{}c@{}}Percentile gap\\ for grid data\end{tabular} & \begin{tabular}[c]{@{}c@{}}NPSQR\\ (pred error)\end{tabular} & \begin{tabular}[c]{@{}c@{}}NPDFSQR\\ (pred error)\end{tabular} \\ \midrule
\multirow{3}{*}{$n=50$} & 5 & 510.31 & 537.40 \\ \cmidrule(l){2-4} 
 & 10 & 524.79 & 568.91 \\ \cmidrule(l){2-4} 
 & 20 & 559.10 & 580.08 \\ \midrule
\multirow{3}{*}{$n=100$} & 5 & 550.09 & 548.82 \\ \cmidrule(l){2-4} 
 & 10 & 578.09 & 571.07\\ \cmidrule(l){2-4} 
 & 20 & 573.44 & 575.94 \\ \midrule
\multirow{3}{*}{$n=200$} & 5 & 497.74 & 541.71 \\ \cmidrule(l){2-4} 
 & 10 & 535.20 & 539.61 \\ \cmidrule(l){2-4} 
 & 20 & 540.32 & 549.71 \\ \bottomrule
\end{tabular}
\caption{(Second Simulation Study (Grid Data)) Prediction Mean Squared Errors using NPSQR and NPDFSQR based on simulation studies with $5,10,20$ distant percentile grids for sample sizes $n=50, 100, 200$.}
\label{tab_grid_2}
\end{table}

\section{Application to Hurricane Data}
\cite{Elsner2008} made an argument that the hurricanes with higher velocities in the North Atlantic basin have got stronger in the last couple of decades. We apply NPSQR method to estimate the simultaneous quantiles of the hurricane velocities in the North Atlantic basin using the hurricane intensity data\footnote{Source \url{http://weather.unisys.com/hurricane/atlantic/}} during the period 1981--2006. First the explanatory variable time is linearly transformed to unit interval such that the years 1981 and 2006 are mapped to 0 and 1 respectively. To transform the hurricane velocities into the unit interval, we assume the hurricane velocities follow the power-Pareto distribution. The power-Pareto density is given by 
\begin{align}
f(y)=\frac{ak(y/\sigma)^{k-1}}{\sigma(1+(y/\sigma)^k)^{(a+1)}} \quad y>0 \nonumber
\end{align}
The distribution function is given by 
\begin{align}
F(y)=1-\frac{1}{(1+(y/\sigma)^k)^a}
\label{eq:pareto_dist_here}
\end{align}
\cite{Tokdar2012} proposed the parameter values as $a=0.45$, $\sigma=52$ and $k=4.9$ in the same context. We transform the hurricane velocities into unit interval using Equation (\ref{eq:pareto_dist_here}).

\begin{figure}
    \centering
    \includegraphics[width=0.7\linewidth]{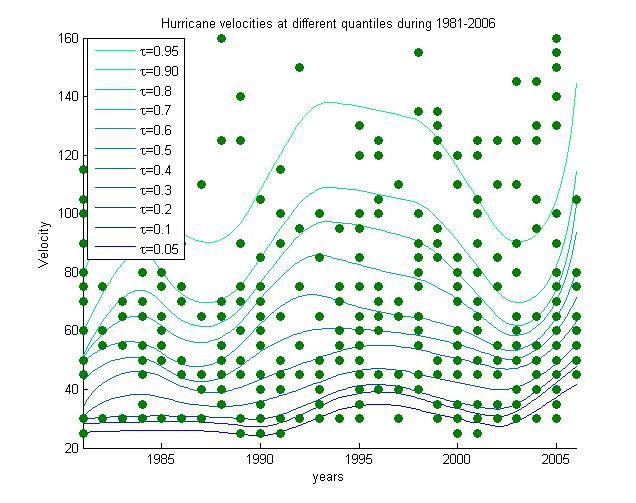}  
  \caption{Simultaneous quantiles of hurricane velocities in North Atlantic region during 1981-2006.}
  \label{figfighurricane} 
\end{figure}

After transforming both $X$ and $Y$ values into unit interval, we use NPSQR method to estimate the simultaneous quantiles of the hurricane wind velocities. Similar to the approach considered in  Section \ref{sec_complete}, we start the Block-Metropolis Hastings algorithm with a warm starting point found using GCDVSMS algorithm. We consider 10000 posterior samples discarding the first 1000 samples as burn-in. The number of equidistant knots to be used for B-spline basis expansion is selected using the AIC. After the quantile curves are estimated, corresponding inverse transformations are performed on the response and the explanatory variables before plotting them. In Figure \ref{figfighurricane} we note that unlike the upper quantiles, the lower quantiles of the hurricane velocities have changed little over the time. We note a periodic pattern in the upper quantiles. Specially this pattern becomes more prominent with increasing values of the quantile. A increasing pattern of the higher quantile curves is noted during the period 1987-1994 and 2002-2005 while a decreasing pattern is prominent during 1994-2002.

\section{Application to US household income data}
Historical tables for US household income data can be found in this site\footnote{Source \url{http://www.census.gov/data/tables/time-series/demo/income-poverty/historical-income-households.html}}. In the data tables, the 20, 40, 60, 80 and 95-th quantiles of the household income in current dollars (accessed 11-25-2016) of all population (combined), Asian, Black, Hispanic, White and White non-Hispanic population have been provided for a few years. Along with that, the total population of each category at each considered year has been given. A snapshot of the data table showing the household income distribution of all population during 2010-2015 is given in Table \ref{snapshot}. For analysis, we transform the years linearly to the the unit interval. For example, for Hispanic population since the data is available during 1972-2015, we transform it linearly to the unit interval such that 1972 and 2015 get mapped to 0 and 1 respectively. To transform the incomes into unit interval, we use a log-linear transformation. First we take logarithms of all income values of all races. After the $\log$ transformation, suppose $a_1$ and $a_2$ denote the smallest and the biggest values. Define $L = a_1-0.01$ and $U = a_2+0.01$. We use the transformation $f(y) = (\log y - U)/(L-U)$ to transform all the incomes to unit interval and the values of $L$ and $U$ come out to be $7.47$ and $12.55$ respectively. After the analysis, inverse transformations are performed before plotting the quantile curves to return to the original scale.
\begin{table}[]
\centering
\centering
\resizebox{\columnwidth}{!}{%
\bgroup
\def\arraystretch{1.2}%
\begin{tabular}{@{}ccccccc@{}}
\toprule
Years & \begin{tabular}[c]{@{}c@{}}Population\\ (thousands)\end{tabular} & \begin{tabular}[c]{@{}c@{}}20 \\ percentile\end{tabular} & \begin{tabular}[c]{@{}c@{}}40 \\ percentile\end{tabular} & \begin{tabular}[c]{@{}c@{}}60 \\ percentile\end{tabular} & \begin{tabular}[c]{@{}c@{}}80 \\ percentile\end{tabular} & \begin{tabular}[c]{@{}c@{}}95 \\ percentile\end{tabular} \\ \midrule
2015 & 125,819 & 22,800 & 43,511 & 72,001 & 117,002 & 214,462 \\
2014 & 124,587 & 21,432 & 41,186 & 68,212 & 112,262 & 206,568 \\
2013 & 123,931 & 21,000 & 41,035 & 67,200 & 110,232 & 205,128 \\
2013 & 122,952 & 20,900 & 40,187 & 65,501 & 105,910 & 196,000 \\
2012 & 122,459 & 20,599 & 39,764 & 64,582 & 104,096 & 191,156 \\
2011 & 121,084 & 20,262 & 38,520 & 62,434 & 101,582 & 186,000 \\
2010 & 119,927 & 20,000 & 38,000 & 61,500 & 100,029 & 180,485 \\ \bottomrule
\end{tabular}
\egroup
}
  \caption{A snapshot of US household income data table showing the income distribution of all population during 2010-2015.}
  \label{snapshot}
\end{table}

It is noted that this data is somewhat different from the data considered for simulation study in Section \ref{sec_grid}. Firstly, the value of quantile grids (i.e., $q_Y(\rho_l)\}_{l=1}^c$) in this data is different for each year (or time-point). For example, as seen in Table \ref{snapshot}, the values of income at different quantile levels in 2015 is different than that of 2014. Secondly, unlike the data considered in Section \ref{sec_grid}, here for each value of $X$ (i.e., time-point), there are multiple observations. The quantile grid considered here is given by $\{\rho_l\}_{l=0}^6$ where $\rho_0=0, \rho_1=0.2, \rho_2=0.4, \rho_3=0.6, \rho_4=0.8, \rho_5=0.95$ and $\rho_6=1$. So if at $n$ time-points $\{X_i\}_{i=1}^n$, the number of observations (i.e., population) are $\{V_i\}_{i=1}^n$ then the likelihood is given by
\begin{align*}
L= & \prod_{i=1}^n(F(q_Y(0.2|X_i)))^{0.2V_i}.(F(q_Y(0.4|X_i)) - F(q_Y(0.2|X_i)))^{0.2V_i}. \\
& (F(q_Y(0.6|X_i)) - F(q_Y(0.4|X_i)))^{0.2V_i}.(F(q_Y(0.8|X_i)) - F(q_Y(0.6|X_i)))^{0.2V_i}. \\
& (F(q_Y(0.95|X_i)) - F(q_Y(0.8|X_i)))^{0.15V_i}.(1-F(q_Y(0.95|X_i)))^{0.05V_i}.
\end{align*}
Clearly, the value of $\{F(q_Y(\rho_l|X_i))\}_{l=0}^6$ can be found using the same technique as used described in Section \ref{sec_grid}.

To estimate the quantile curves we use the NPSQR method. We start the Block Metropolis-Hastings MCMC algorithm with warm starting point found using the GCDVSMS algorithm. Unlike all the previous studies, this data represent the population, not the sample. Hereby, instead of choosing the optimal number of knots for fitting the B-spline, we can fix their values anywhere depending on desired smoothness level. We set $p_1=p_2=5$ for the whole analysis.

\begin{figure}[] 
   \begin{subfigure}[b]{0.5\linewidth}
    \centering
    \includegraphics[width=0.99\linewidth]{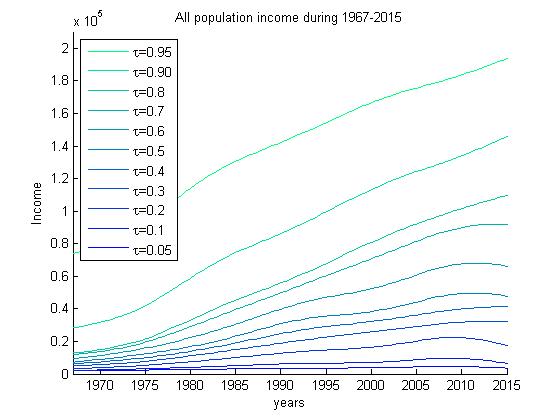} 
    \caption{All population}  
    \label{income_1} 
  \end{subfigure} 
   \begin{subfigure}[b]{0.5\linewidth}
    \centering
    \includegraphics[width=0.99\linewidth]{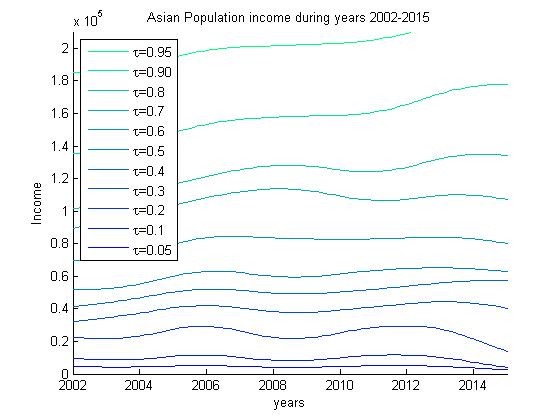} 
   \caption{Asian}  
    \label{income_2} 
  \end{subfigure}
\begin{subfigure}[b]{0.5\linewidth}
    \centering
    \includegraphics[width=0.99\linewidth]{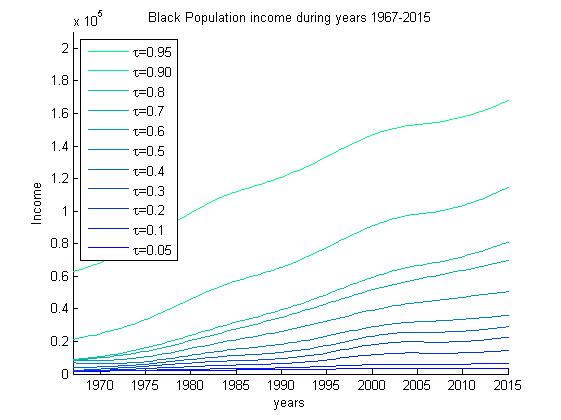} 
    \caption{Black} 
    \label{income_3} 
  \end{subfigure}
  \begin{subfigure}[b]{0.5\linewidth}
    \centering
    \includegraphics[width=0.99\linewidth]{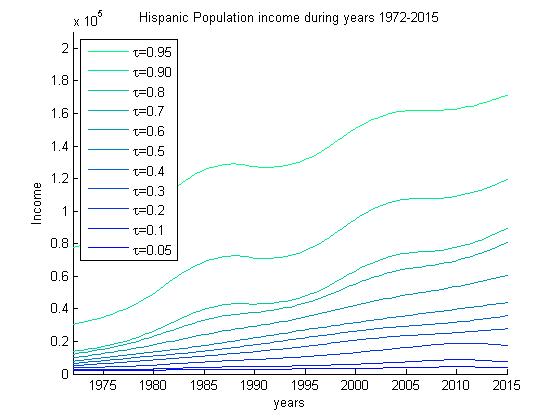} 
    \caption{Hispanic} 
    \label{income_4} 
  \end{subfigure} \\
  \begin{subfigure}[b]{0.5\linewidth}
    \centering
    \includegraphics[width=0.99\linewidth]{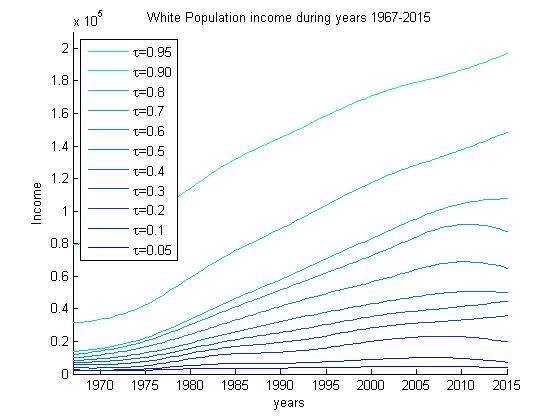} 
    \caption{White} 
    \label{income_5} 
  \end{subfigure}
    \begin{subfigure}[b]{0.5\linewidth}
    \centering
    \includegraphics[width=0.99\linewidth]{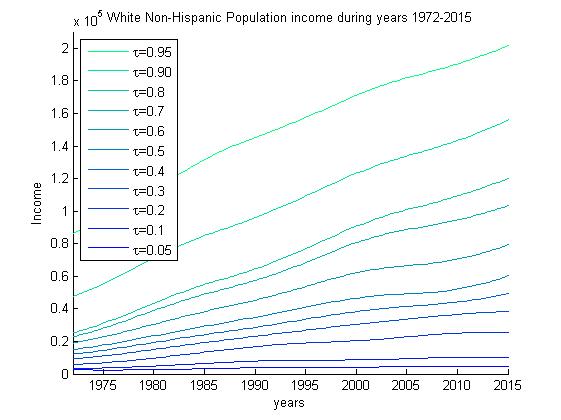} 
    \caption{White non-Hispanic}  
    \label{income_6} 
  \end{subfigure} \\
  \caption{Estimated simultaneous quantiles of household income of (a) All population (1967-2015) (b) Asian (2002-2015) (c) Black (1967-2015) (d) Hispanic (1972-2015) (e) White (1967-2015) and (f) White Non-Hispanic (1972-2015) population at $\tau = \{0.05, 0.10, 0.20, \ldots, 0.90, 0.95\}$ for $n=100$ using NPSQR.}
  \label{INCOME} 
\end{figure}

In Figure \ref{INCOME} we plot the estimated simultaneous quantiles of household income of all population (during 1967-2015), Asian (during 2002-2015), Black (during 1967-2015), Hispanic (during 1972-2015), White (during 1967-2015) and White non-Hispanic (during 1972-2015) population. Irrespective of the races, it is noted that the higher quantile curves increase at higher rates while the lower quantile curves are roughly constant over the years. Household incomes seem to be more evenly distributed among Asian people compared with other races. Among other races White and White non-Hispanic population have lesser gaps across different quantile levels of income distribution compared with that of Black and Hispanic population. The estimated quantile levels of household income of Asian population is greater than other populations. It should be also noted that the gaps between the quantile levels are more or less increasing over time.

\section{Conclusion}
In this paper two novel methods for non-parametric simultaneous quantile regression methods have been proposed. In the first method, the quantile function is estimated non-parametrically using tensor products of quadratic B-splines basis expansion and in the second method the distribution function is estimated by a non-parametric approach using tensor product of quadratic B-splines basis expansion. These methods have been further developed for estimating the quantiles for the quantile grid data. We consider the Block Metropolis-Hastings MCMC algorithm to estimate the coefficients of the B-spline basis functions. Before initializing the MCMC, the Maximum Likelihood Estimator (MLE) is evaluated using GCDVSMS algorithm and it is used as the starting point. The optimal number of knots  for the B-spline basis expansion is selected using the AIC criterion. Unlike the existing popular methods of non-linear quantile regression, e.g., local linear and local spline quantile regression, the monotonicity of the quantile curves are maintained using the proposed methods. In the simulation studies it is shown that both of the proposed methods generally perform better than LLQR and LSQR in terms of the PMSE. It is also observed that for the quantile grid data, NPSQR performs slightly better than NPDFSQR in terms of PMSE.

The NPSQR method has been used to analyze the hurricane intensity data of North Atlantic region for the years 1981-2006. A periodic nature is noted at higher quantile levels while estimated lower quantile curves are relatively stable and with respect to time. NPSQR method has been also used to analyze the historical household income data of different races in US given in the form of quantile grids. It is noted that the higher quantile curves are increasing generally at higher rates than the lower quantile curves. The differences between the household income levels tend to increase over time.

\end{document}